\documentclass[conference]{IEEEtran} 
\usepackage{amsmath}
\usepackage{amssymb}
\usepackage{tabularx}
\usepackage{mathtools}
\usepackage{graphicx}
\usepackage{optidef}
\usepackage{amsfonts}
\usepackage{algorithm}
\usepackage{algorithmic}
\usepackage{caption}
\usepackage{subcaption}
\usepackage{amsthm}

\usepackage{xcolor}
\usepackage{breqn}
\usepackage{cuted}
\usepackage[top=0.75in, left=0.64in, right=0.64in, bottom=1.04in, letterpaper]{geometry}

\ifCLASSINFOpdf
\else
\fi
\DeclareUnicodeCharacter{2212}{-}

\begin{document}
%
\title{DRL-based Joint Resource Scheduling of eMBB and URLLC in O-RAN} 
%
%
%
\author{
     Rana M. Sohaib\IEEEauthorrefmark{2}, \IEEEauthorblockN{Syed Tariq Shah\IEEEauthorrefmark{1}, Oluwakayode Onireti\IEEEauthorrefmark{2}, Yusuf Sambo\IEEEauthorrefmark{2}, Qammer H.
     Abbasi\IEEEauthorrefmark{2}, M. A. Imran\IEEEauthorrefmark{2}}
    \IEEEauthorblockA{
    \IEEEauthorblockA{\IEEEauthorrefmark{2}James Watt School of Engineering, University of Glasgow, Glasgow, UK\\ Emails: \{RanaMuhammad.Sohaib, Oluwakayode.Onireti, Yusuf.Sambo, Qammer.Abbasi, Muhammad.Imran\}@glasgow.ac.uk}
    \IEEEauthorrefmark{1}School of Computer Science \& Electronic Engineering, University of Essex, Colchester, UK\\ Email: Syed.Shah@essex.ac.uk}
}



\maketitle

\begin{abstract}
This work addresses resource allocation challenges in multi-cell wireless systems catering to enhanced Mobile Broadband (eMBB) and Ultra-Reliable Low Latency Communications (URLLC) users. We present a distributed learning framework tailored to O-RAN network architectures. Leveraging a Thompson sampling-based Deep Reinforcement Learning (DRL) algorithm, our approach provides real-time resource allocation decisions, aligning with evolving network structures. The proposed approach facilitates online decision-making for resource allocation by deploying trained execution agents at Near-Real Time Radio Access Network Intelligent Controllers (Near-RT RICs) located at network edges. Simulation results demonstrate the algorithm's effectiveness in meeting Quality of Service (QoS) requirements for both eMBB and URLLC users, offering insights into optimising resource utilisation in dynamic wireless environments. 
\end{abstract}


\IEEEpeerreviewmaketitle

\section{Introduction}
Traditional radio access networks (RAN) often rely on a single-vendor model, presenting a monolithic, black-box solution that hinders flexibility and innovation. In contrast, open RAN (O-RAN) disrupts this paradigm by separating hardware from software and promoting open interfaces \cite{b1}. This allows network operators to customise their infrastructure according to the requirements.
The architecture of O-RAN incorporates several foundational elements, namely the Radio Unit (RU), Distributed Unit (DU), Central Unit (CU), and RAN Intelligent Controller (RIC). 
RIC is an innovative network component that introduces new services and optimisation capabilities in the network.
More specifically, RIC encompasses both near real-time and non-real-time variants, fostering programmable-based functions that significantly enhance network flexibility and efficiency. By integrating embedded AI capabilities, RIC can dynamically adapt radio resource \cite{STSJWCN}, mobility, and spectrum management operations, which includes admission control, radio resource allocation and scheduling \cite{STSRA}, power allocation, and radio link management, to meet the specific requirements of applications. This adaptability proves especially crucial in Beyond 5G (B5G) networks, which cater to a diverse range of vertical industries, by ensuring network operations are finely tuned to serve varying demands efficiently \cite{Rana2}.

In the context of O-RAN in B5G networks, two pivotal service categories are enhanced Mobile Broadband (eMBB) and Ultra-Reliable low latency communications (URLLC). The eMBB targets high data rate applications, aiming for improved data rates and reliability, with a focus on supporting a packet error rate (PER) target of $10^{-3}$. URLLC, in contrast, is engineered for critical applications requiring immediate data transmission with minimal delay, such as autonomous driving and remote surgery. It prioritises ultra-high reliability (PER around $10^{-5}$) and low latency, necessitating significant physical and MAC layer adjustments in 5G NR to increase throughput capacity and reduce transmission time intervals, control signalling overhead and outage probability \cite{Rana3}. Accommodating these requirements is crucial for O-RAN's adaptability and performance in diverse B5G applications.
The coexistence of eMBB and URLLC services within the same network infrastructure necessitates innovative radio resource management (RRM) strategies to satisfy the contrasting requirements of high throughput for eMBB and low latency and high reliability for URLLC \cite{RanaS11}. 
Conventional optimisation approaches face limitations in providing real-time solutions for the coexistence of eMBB and URLLC \cite{rananew}. 
Deep Reinforcement Learning (DRL) is essential in this context as it enables systems to learn and adapt in real time, leveraging experience to make decisions in dynamic environments. 
The authors in \cite{RanaR1} present a dual-strategy approach integrating risk-sensitive optimisation and DRL to optimise resource allocation for eMBB and URLLC coexistence in 5G. Their proposed approach improves the reliability of URLLC while protecting the eMBB reliability. The superposition scheme's efficacy for facilitating uplink communication among eMBB, mMTC, and URLLC service classes is addressed in \cite{RanaR3}, where the authors highlight the advantages of employing superposition in this context.
Existing DRL-based solutions for eMBB and URLLC co-existence employed the $\epsilon$-greedy approach for action policy, primarily focused on exploitation by consistently selecting the action estimated to yield the highest potential reward. This approach may result in sub-optimal solutions or being trapped in local optima. 
Unlike the above-mentioned studies, which mainly focus on eMBB and URLLC co-existence in a generic ideal RAN scenario, this paper focuses on the non-ideal front haul (FH) scenario. In this paper, we employ the Thompson sampling-based DRL approach to balance exploration and exploitation, addressing the co-existence challenges between eMBB and URLLC in O-RAN. 
\section{System Model}
\begin{figure}[t]
     \centering
\includegraphics[width=0.45\textwidth]{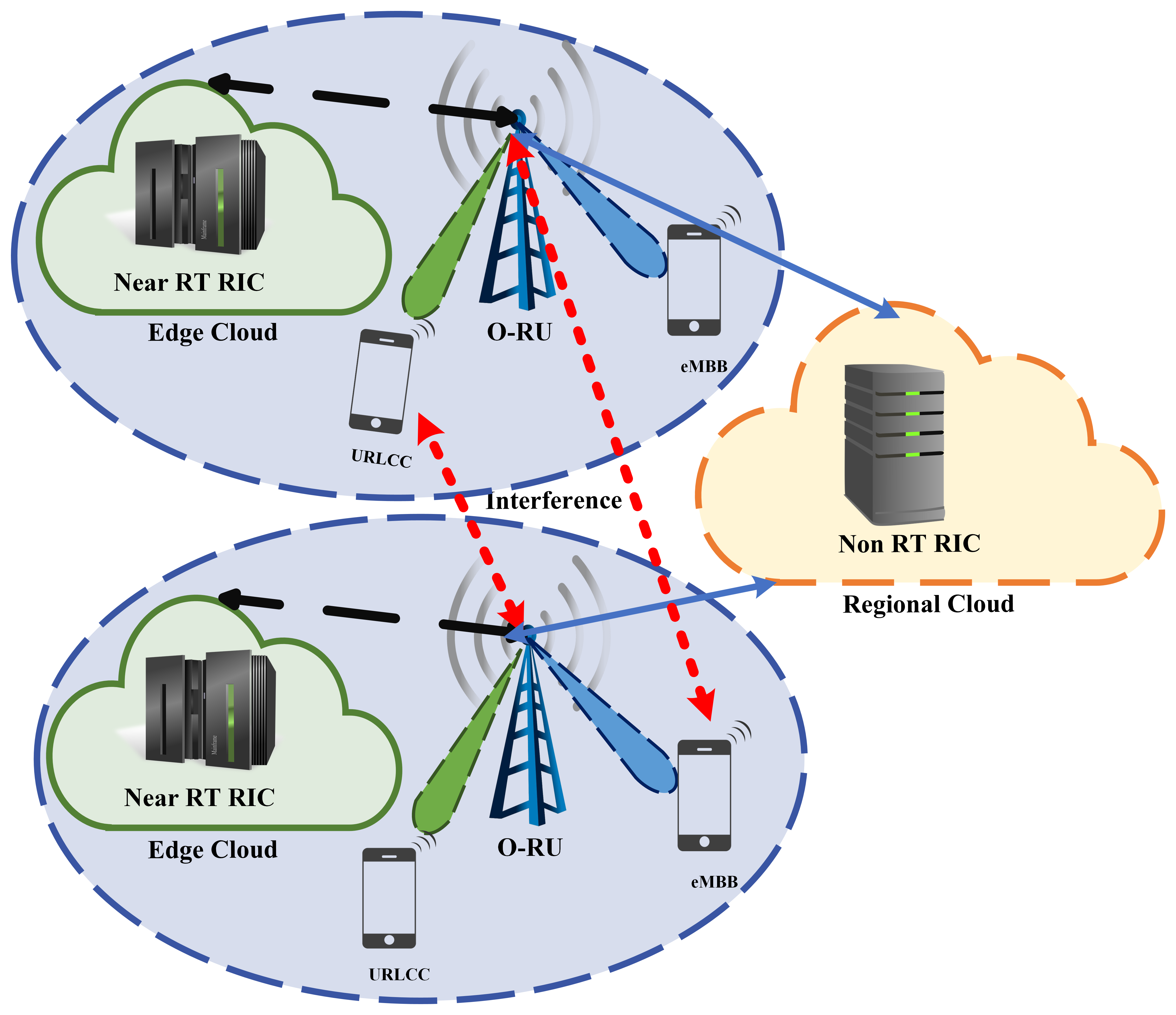}
\caption{Illustration of System model of O-RAN}
\label{fig:fig21}
\end{figure}

We model a wireless network that delivers two distinct services, namely, eMBB and URLLC, as illustrated in Fig. \ref{fig:fig21}, where we strategically placed several edge cloud servers at the near-RT-RIC, establishing connections with a regional cloud server at non-RT-RIC. According to \cite{b1}, Non-RT RIC is situated at the regional cloud server, while the Near-RT RICs are implemented on the edge cloud servers.  In the considered model, we have multiple small cells, and set of all BS is defined as $\mathcal{K}=\left\{1, . . ., K \right\}$, where a BS $k\in K$ covers a set of eMBB users $\mathcal{V}_{k}^{e}=\left\{1, . . ., V_{k}^{e} \right\}$, and URLLC users $\mathcal{V}_{k}^{u}=\left\{1, . . ., V_{k}^{u} \right\}$ present in the network. Moreover, each BS is linked with a single-edge cloud server. The radio resources in 5G-NR can be depicted in both the frequency and time domains. These domains are further subdivided into a set of $M$ radio resources, commonly known as RB. Each RB is characterised by a bandwidth denoted as $B$. Every time slot is subdivided into $L$ mini-slots.
In general, the eMBB service extends over several TTIs to improve spectral efficiency (SE). However, due to the strict latency requirements, the incoming URLLC traffic cannot be delayed.
Puncturing eMBB slots are employed to address URLLC's stringent latency and reliability demands. By puncturing eMBB slots, dedicated resources can be allocated to URLLC traffic, ensuring timely and reliable delivery of critical information. The URLLC service is scheduled with a brief TTI of 0.5 ms, while a longer 1 ms duration is allocated for the eMBB service.
The immediate scheduling of URLLC transmission, which entails disrupting eMBB traffic, can exert a notable influence on the system's capacity and reliability. This, in turn, may result in a decline in the eMBB service's performance. 
Therefore, an appropriate framework is required to fulfil the QoS requirements.\\
\textbf{eMBB rate:}
The immediate scheduling of URLLC transmission can degrade the rate of the eMBB service. 
A puncturing decision variable is introduced as follows:
\begin{align}
    \eta_{m,l}^{k,v}(t)=
    \begin{cases}
      1, &\text{if the $l^{th}$ mini-slot is punctured by the} \\
      & \text{$v^{th}$ URLLC user, $\forall$ $m \in \mathcal{M}$}, k\in\mathcal{K} \\
      0, & \text{otherwise}.
    \end{cases}
\end{align}
The signal-to-noise-interference-ratio (SINR) of the
eMBB user $v$ can be determined as follows
\begin{equation}
    \chi_{k,m}^{e,v}(t)\!=\!\frac{p_{k,m}^{e,v}(t)g_{k,m}^{e,v}(t)}{\sum\limits_{\substack{k'\in\mathcal{K} \\ k'\neq k}}\underbrace{p_{k',m}^{e,v}(t)g_{k',m}^{e,v}(t)}_{\text{eMBB interference}}+\!\!\sum\limits_{\substack{k'\in\mathcal{K} \\ k'\neq k}}\underbrace{p_{k',m}^{u,v}(t)g_{k',m}^{u,v}(t)}_{\text{URLLC interference}}+\sigma^{2}},
\end{equation}
where $p_{k,m}^{e,v}(t)$, and $g_{k,m}^{e,v}(t)$ refers to the transmitted power and channel gain, respectively, of eMBB user $v$ of BS $k$ over RB $m$, and $\sigma^{2}$ denotes the noise power. The achievable rate of an eMBB
user $v$ of BS $k$ on RB $m$ at time slot $t$ can be computed as
\begin{align}
    r_{k,m}^{e,v}(t)=B\left(1-\frac{\sum_{l=1}^{L}\eta_{m,l}^{k,v}(t)}{L} \right)\log_{2}\big(1+\chi_{k,m}^{e,v}(t) \big),
\end{align}
where the expression $\frac{\sum_{l=1}^{L}\eta_{m,l}^{k,v}(t)}{L}$ indicates the degradation of eMBB rate caused by puncturing. We make the assumption that each BS reserves an RB for a single user. We define the RB allocation binary decision variable as $(\beta_{v,m}^{k}(t))$, that takes the value of 1 if RB $(m)$ of BS $(k)$ is assigned to the eMBB user $(v)$, for all $(k)$ in the set $(\mathcal{K})$. It is set to 0 in all other instances. 
Thus, the total sum rate obtained by the eMBB user $v$ can be computed as
\begin{align}
     r_{k,v}^{e}(t)=\sum\limits_{m\in\mathcal{M}}\beta_{v,m}^{k}(t)r_{k,m}^{e,v}(t).
\end{align}
\textbf{URLLC rate:}
To ensure minimal transmission delay, it is necessary to limit the blocklength in URLLC. Shannon's capacity theorem is relevant when dealing with an infinite blocklength. A study in \cite{bs2} delves into analysing resource management challenges specific to URLLC services, considering the achievable data rate within the finite blocklength regime. The achievable rate of URLLC for finite blocklength can be computed as
\begin{align}
     r_{k,m}^{u,v}(t)=&\sum\limits_{m\in\mathcal{M}}B_{m}\Big(\frac{\sum_{l=1}^{L}\eta_{m,l}^{k,v}(t)}{L} \Big)\Bigg[\log_{2}\big(1+\chi_{k,m}^{u,v}(t) \big)\\ 
     &-\sqrt{\!\frac{\!Y_{k,m}^{u,v}}{\!W_{k,m}^{u,v}(t)}}.Q^{\!-1}(\!x)\Bigg],\nonumber
\end{align}
where $W_{k,m}^{u,v}(t)$ refers to the number of symbols, and
$\chi_{k,m}^{u,v}(t)$ denote the SINR of URLLC user, formulated as
\begin{equation}
    \chi_{k,m}^{u,v}(t)=\frac{p_{k,m}^{u,v}(t)g_{k,m}^{u,v}(t)}{\sum\limits_{\substack{k'\in\mathcal{K} \\ k'\neq k}}\underbrace{p_{k',m}^{u,v}(t)g_{k',m}^{u,v}(t)}_{\text{URLLC interference}}+\!\!\sum\limits_{\substack{k'\in\mathcal{K} \\ k'\neq k}}\underbrace{p_{k',m}^{e,v}(t)g_{k',m}^{e,v}(t)}_{\text{eMBB interference}}+\sigma^{2}}.
\end{equation}
Here, $Y_{k,m}^{u,v}=1-\frac{1}{\left(1+\chi_{k,m}^{u,v}(t)\right)^{2}}$ refers to the dispersion of the channel. 
\begin{figure}[t!]
     \centering
\includegraphics[width=0.45\textwidth]{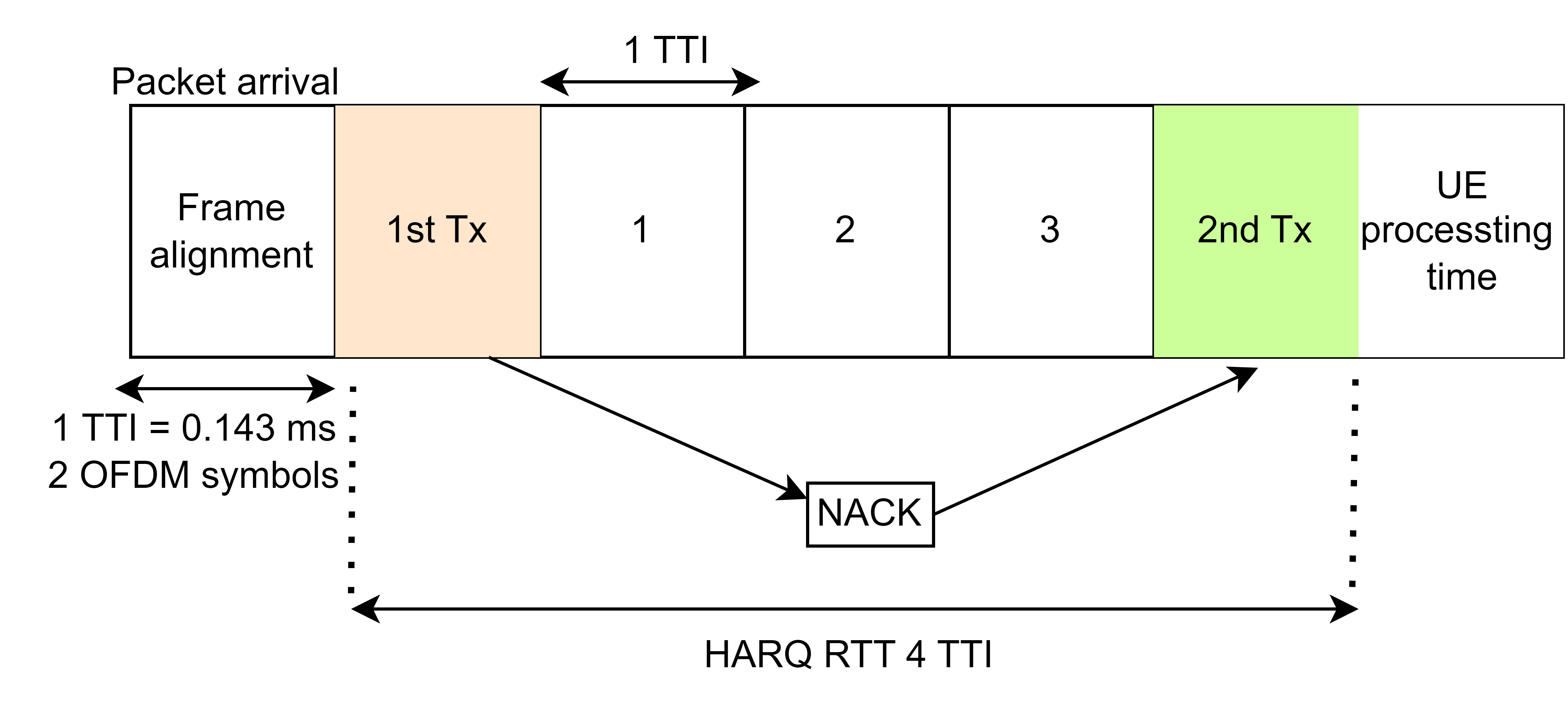}
\caption{URLLC 2 symbols HARQ RTT}
\label{fig:figharq}
\end{figure}
\subsection{URLLC frame structure}
The URLLC frame structure in 5G is designed to meet stringent latency requirements for mission-critical applications. The 3GPP has acknowledged the diverse needs of users and adopted different TTI durations, such as 0.5 ms slots composed of 7 OFDM symbols or 0.143 ms mini-slots composed of 2 OFDM symbols, are considered to provide greater flexibility and adaptability to various use cases \cite{b3}. In this work, we consider a mini-slot comprised of 2 OFDM symbols. 
Since we consider non-ideal FH, it's crucial to address the challenge posed by hybrid automatic repeat request (HARQ) round-trip time (RTT) and the associated latency target of 1 ms for URLLC. The standard $8\times0.143$ ms HARQ RTT exceeds the specified 1 ms latency target, which may impact reliability. To mitigate this, the study explores the reduction of HARQ RTT to 4 TTIs as shown in Fig. \ref{fig:figharq}, allowing room for one HARQ retransmission. Considering one HARQ retransmission, the maximum allowable latency is reduced to 0.86 ms (6x0.143 ms), which meets the strict latency demands of URLLC service. This calculation does not include queuing delay. 
\section{Problem Formulation}
Organising URLLC traffic alongside eMBB users experiencing varying channel quality imposes additional strain on eMBB transmissions. 
We assume that the URLLC users generate small packet fragments. The rate at which packets arrive in mini-slot $l$, where $l$ belongs to the set $\mathcal{L} = \{1, ..., l, ..., L\}$, during TTI $t$, follows a Poisson point process (PPP) distribution and is represented by $\phi_{l}(t)$.
Moreover,  the total number of URLLC packets received during the TTI $t$ can be calculated as $\phi(t)=\sum_{l\in \mathcal{L}}\phi_{l}(t)$. 
Based on $\phi(t)$, the reliability of URLLC service can be estimated as
\begin{align} \label{REL}
   \Pr\left[ r_{k,m}^{u,v}(t)\leq \rho\phi(t)\right]\leq \Psi_{u},\hspace{1cm}\forall k\in K
\end{align}
where $\rho$ denotes the packet size associated with the URLLC service. 
More specifically, the expression states that the probability of the achievable data rate of a URLLC user at a given time slot $t$ is less than or equal to the product of URLLC packet size. The total number of arrived URLLC packets should be less than or equal to a small positive value, denoted $\Psi_{u}$. 
In practical terms, this ensures that the system can reliably handle the communication requirements of the URLLC traffic and also ensures that the probability of failure (not meeting the specified data rate constraint) remains below a predefined limit. 
As mentioned earlier, the proposed system model considers non-ideal FH, and in order to incorporate the round-trip delay of HARQ retransmission into the outage probability of URLLC service, we have to modify the reliability equation given at (\ref{REL}).
In other words, the HARQ retransmissions introduce an additional delay, and the outage probability should consider the reliability of both the initial transmission and potential retransmissions.
Therefore, the updated reliability considering non-ideal FH and HARQ can be expressed as 
\begin{align}\label{eq:out}
   \Pr\left[r_{k,m}^{u,v}(t+\Lambda_{RTT})\leq \rho\phi(t)\right]\leq \Psi_{u},\hspace{0.75cm}\forall k\in K,
\end{align}
where $\Lambda_{RTT}$  denotes the round-trip delay of HARQ retransmission in TTI units. This modification reflects the fact that we are considering the achievable data rates for URLLC users at time $(t+\Lambda_{RTT})$ instead of time $t$. 
The modification to $r_{k,m}^{u,v}(t+\Lambda_{RTT})$ is meant to account for the time at which we assess the reliability of the system.
The success probability for each HARQ transmission can be modelled based on the BER \cite{bhar1, bhar2}. 
At time $(t+\Lambda_{RTT})$, we estimate the probability of successful transmission by considering the CDF of the BER. A Bernoulli experiment is run to determine whether the Transport Block (TB) transmitted to each user is successful or not based on the estimated success probabilities. This modelling approach allows us to evaluate the impact of delays on the success or failure of HARQ transmissions and, subsequently, on the overall data rate for URLLC services.
Consequently, we formulate an objective function aimed at enhancing the allocation of eMBB resource blocks, optimising transmission power for eMBB users, and refining the scheduling strategy for URLLC users in the following manner.
\begin{subequations}\label{eq:op}
\begin{align}
\textbf{P}:\max_{\beta, P, \eta}&\left\{\sum\limits_{v\in\mathcal{V}_{k}^{e}}r_{k,v}^{e}\right\} \label{eq:obf} \\
\textrm{subject to}&\sum_{v\in \mathcal{V}_{k}^{e}}\!\beta_{v,m}^{k}(t)\leq1,\hspace{1.25cm}\forall m\in\mathcal{M}, k\in\mathcal{K} \label{eq:c1}\\
&\sum_{v\in \mathcal{V}_{k}^{u}}\!\eta_{m,l}^{k,v}(t)\leq1,\hspace{1.25cm}\forall m\in\mathcal{M}, k\in\mathcal{K}\label{eq:c2}\\
&\sum_{l\in \mathcal{L}}\!\eta_{m,l}^{k,v}(t)\leq L,\hspace{1.05cm}\forall m\in\mathcal{M}, k\in\mathcal{K}\label{eq:c3}\\
&\Pr\left[ r_{k,m}^{u,v}(t+\Lambda_{RTT})\leq \rho\phi(t)\right]\leq \Psi_{u},\hspace{0.10cm}\forall k\in K\label{eq:c4}\\
&\sum_{v\in\mathcal{V}_{k}^{e}}\sum_{m\in\mathcal{M}}p_{k,m}^{e,v}(t)\leq P_{max},\hspace{0.5cm}\forall k\in\mathcal{K}\label{eq:c5}\\
&p_{k,m}^{e,v}(t)\geq 0, \hspace{1.55cm}\forall v\in\mathcal{V}^{e}, m\in\mathcal{M} \label{eq:c6}\\
&\beta_{v,m}^{k}(t)\in\{0,1\}, \hspace{1cm}\forall v\in\mathcal{V}^{e}, m\in\mathcal{M}\label{eq:c7}\\
&\eta_{m,l}^{k,v}(t)\in\{0,1\}, \hspace{1cm} \forall v\in\mathcal{V}^{u}, m\in\mathcal{M}\label{eq:c8}
\end{align}
\end{subequations}
where (\ref{eq:obf}) refers to the objective where we look to maximize the rate of eMBB service. Constraint (\ref{eq:c1}) defines the eMBB RB allocation constraint, ensuring that only one user is linked to a RB. 
Whereas (\ref{eq:c2}) ensures that no more than one URLLC user can puncture a specific mini-slot on a particular RB at a given time slot. Constraint (\ref{eq:c3}) defines that the punctured mini-slots should be less than or equal to the total number of mini-slots. Constraint (\ref{eq:c4}) defines the URLLC reliability constraint. Constraints (\ref{eq:c5}) and (\ref{eq:c6}) refer to the eMBB power allocation constraints. Likewise, the constraints (\ref{eq:c7}) and (\ref{eq:c8}) specify the limitations on resource allocation visibility.
\section{Proposed Decentralized DRL-based Framework}
The DRL agents are positioned at the edges of the network, while a core training module operates from the regional cloud server for simplicity of implementation and enhanced stability. The central cloud server conducts training globally, utilising experiences accumulated from all agents. This method facilitates quicker convergence and improves overall performance. Despite sharing common learning parameters from the central cloud server, each agent independently takes decisions, with no awareness of the decisions made by others. 
Solving the mixed-integer programming challenge presented in (\ref{eq:op}) is typically classified as NP-hard. Furthermore, the scheduling parameter for URLLC is intricately linked with the RB and power allocation parameters, thereby increasing the intricacy of the optimisation challenge. We present a DRL-based framework to address the optimisation problem presented in (\ref{eq:op}), where we model it as a Markov decision process (MDP) for $K$ agents.  
The agent selects an action from its action space based on the observations associated with the state, and the action is guided by the policy $\pi$. 
\subsubsection*{State space}
The set of state space can be defined as $S=\left\{S_1, S_2, . . . , S_k, . . ., S_K \right\}$. 
We assume that each edge cloud server acting as an agent exclusively gets its individual state, specifically the information of users within the same cell. This approach is adopted to minimise the overhead resulting from information exchange across cells.  It can be represented as $s_{k}(t)=\left\{g_{k}^{e}(t), g_{k}^{u}(t), \phi_{k}(t), V_{k}^{e}, V_{k}^{u}\right\}$, where it comprises the channel information of both eMBB and URLLC users, along with the traffic information at time slot $t$.  
\subsubsection*{Action}
The set of action space can be defined as $A=\left\{A_1, A_2, . . . , A_k, . . ., A_K \right\}$. Each agent takes action on selecting eMBB RB $\beta$, eMBB power allocation $P$, and URLLC scheduling $\eta$. 
\subsubsection*{Reward}
We design the global reward function by considering the requirements of URLLC service. The reward function can be formulated as
\begin{align}\label{eq:rewa}
r(t)\!\!=&\big(\!\!\overbrace{{\sum\limits_{v\in\mathcal{V}_{k}^{e}}\!r_{k,v}^{e}}}^{I}\!\big)\!-\Phi(t)\big(\!\!\overbrace{\sum\limits_{v\in\mathcal{V}_{k}^{u}}\!r_{k,m}^{u,v}(t+\Lambda_{RTT})-\rho\phi(t) }^{II}\!\big) 
\end{align}
where we introduce the time-varying weight coefficient $\Phi(t)$ to ensure the URLLC reliability constraint. It can be updated as 
\begin{align}
    \Phi(t+1)=\max\left\{\Phi(t)+\Psi(t)-\Psi_{u},0\right\},
\end{align}
where $\Psi(t)$ refers to the achieved outage probability as mentioned in (\ref{eq:out}). In the first part, we aim to maximise the rate of eMBB users, while the second part represents the constraint for URLLC. The objective of the agent is to choose an optimal policy ($\pi$) by balancing the trade-off between low outage probability and high eMBB data rate.
\subsection{Thompson sampling-based DDPG framework}
The DDPG agent consists of two main components: an actor and a critic. The actor is responsible for defining a policy function, which maps states to corresponding actions. The policy, denoted as $\pi = \pi_{a}^{L}$, can be characterised by the network state observed by the agent. The agent then performs actions corresponding to the number of punctured mini-slots $L$ from each allocated RB. Subsequently, the reward is computed by the agent using equation (\ref{eq:rewa}), considering the decisions made. The updated state information of the network is then provided to the agent. The objective of the actor is to acquire a strategy that maximises the cumulative discounted reward. It can be denoted as $J=\sum_{t=1}^{T}\mu r(t)\big(s(t), a(t)\big)$, where
$\mu$ refers to the discount factor. To assess the effectiveness of the actor's chosen actions, the critic employs a state-action value function denoted as $Q(s(t), a(t))$. This function calculates the cumulative return over the long term by considering the immediate reward $(r(t)(s(t), a(t)))$ and the estimated future returns $(\pi Q(s(t+1), a(t+1)))$ resulting from taking action at in-state $s(t)$ according to the current policy. It utilises a Deep DNN with parameters $\Theta^Q$, denoted as $Q(s(t), a(t) |\Theta^Q)$, to compute the Q-value for a given state-action pair $(s(t), a(t))$. 
To ensure a stable learning process, both the critic and actor utilise duplicates of their DNNs. The critic employs Q-networks, denoted as $\acute{Q}(s(t+1), a(t+1)|\Theta^{\acute{Q}})$, parameterized by $\Theta^{\acute{Q}}$, while the actor utilizes a policy network $\acute{\pi}(s(t+1)|\Theta^{\acute{\pi}})$, parameterized by $\Theta^{\acute{\pi}}$. The critic undergoes training to enhance the optimisation of $\Theta^{Q}$ and $\Theta^{\acute{Q}}$, ensuring accurate computation of the long-term return. The update of $\Theta^{Q}$ using gradient descent involves minimising the loss function as follows
\begin{align}
    \Upsilon(\Theta^{Q})=(Q( s(t), a(t) | \Theta^{Q}) - \zeta(t) )^{2},
\end{align}
where $\zeta(t)= r(s(t), a(t)) + \mu\acute{Q}(s(t+1), a(t+1) |\Theta^{\acute{Q}} )$ refers to the target action-value. The parameter $\Theta^{\acute{Q}}$ is updated as follows $\Theta^{\acute{Q}}=\varkappa\Theta^{Q}+(1-\varkappa)\Theta^{\acute{Q}}$. The actor undergoes training to discover the most effective policy that maximizes the objective function $J$, essentially identifying the optimal values for $\Theta^{\pi}$ and $\Theta^{\acute{\pi}}$. The parameter $\Theta^{\pi}$ is updated using a gradient ascent algorithm along with the corresponding gradient as follows
\vspace{-0.2cm}\begin{align}
    \nabla_{\Theta^{\pi}}J=\nabla_{\pi}Q(s(t), a(t) | \Theta^{Q})\nabla_{\Theta^{\pi}}\pi\left(s(t)|\Theta^{\pi} \right), 
\end{align}
where $\nabla_{\pi}Q\big(s(t), a(t) | \Theta^{Q}\big)$ denotes the policy evaluation.
In this paper, we use Thompson sampling to balance the exploration and exploitation. When an edge agent selects a hypothesis before interacting with the environment, uncertainty arises. Hence, achieving a balance between exploration and exploitation becomes imperative to identify the optimal policy. 
Each hypothesis, associated with parameter $\theta$, undergoes training on a relevant subset $D$ to shape the distribution over the posterior knowledge $\acute{P}(\theta|D)$. Approximating $\acute{P}(\theta|D)$ entails capturing the posterior distribution, a challenge arising from the difficulty of precisely determining values. Instead, a variety of hypotheses is extracted, and their distribution is dictated by the prior distribution $P(\theta)$.
We consider a set of actors $A\in\mathcal{A}$, each defined by specific parameters $\theta_{i}$, where $i\in \mathcal{A}$. These parameters are sampled from the prior distribution $P(\theta)$. The hypothesis in question encapsulates this ensemble of actors, giving rise to the distribution $\acute{P}(\theta|D)$, which is influenced by the original distribution $P(\theta)$. To maintain the distribution, each actor undergoes personalised updates contingent on its specific sub-sample, as indicated by the mask value derived from a Bernoulli distribution \cite{b4}. This binary mask value serves to identify whether a given sample is part of the actor's subset or not. The implementation of Thompson sampling entails a stochastic selection from the ensemble of actors. This random selection process informs the formation of a hypothesis aligned with the posterior distribution. This adaptive strategy plays a pivotal role in striking a nuanced balance between exploration and exploitation, contributing to the formulation of an optimal policy for the eMBB resource block, power allocation and URLLC scheduling. Therefore, the target function is
updated as follows
\begin{align}
 \!\!\!   \zeta(t)\!=\! r\!\left(s(t), a(t)\right)\! +\!\mu\max_{i\in\mathcal{A}}\!\bigg[\acute{Q}_{i}\!\!\left(\!s(t+1), a(t+1) |\Theta^{\acute{Q}} \right)\!\!\bigg].
\end{align}
A neural network model undergoes training offline at the Non-RT RIC. This controller is situated in the regional cloud server and utilizes data gathered from all agents. After the training process, the model is communicated to the DRL agents located at the Near-RT RICs on edge cloud servers. The training objective for the global model is to maximize the predefined global reward function outlined in (\ref{eq:rewa}).
\begin{table}[htbp]
\caption{Simulation Parameters} 
\centering
\label{tab:1}
\begin{tabularx}{\columnwidth}{|X|X|}
\hline
\textbf{Parameter} & \textbf{Value} \\ 
\hline\hline
Frame duration & 10 ms \\ 
\hline
No. of mini-slots in each TTI & 7 \\ 
\hline
sub-carrier spacing & 15 Khz \\ 
\hline
No. of OFDM symbols/TTI & 14 \\ 
\hline
OFDM symbols/mini-slot & 2 \\ 
\hline
Bandwidth & 20 MHz \\ 
\hline
URLLC packets length & 32 Bytes  \\ 
\hline
RB Bandwidth & 180 kHz \\ 
\hline
Transmit power & 38 dBm \\ 
\hline
Pathloss Model & $120.8 + 37.5 \log_{10}(d)$ \\ 
\hline
Actor learning rate & $10^{-5}$ \\ 
\hline
Critic learning rate & $10^{-3}$ \\ 
\hline
\end{tabularx}
\end{table}
\begin{figure}[htbp]
\vspace*{0.02in}
\includegraphics[width=0.45\textwidth]{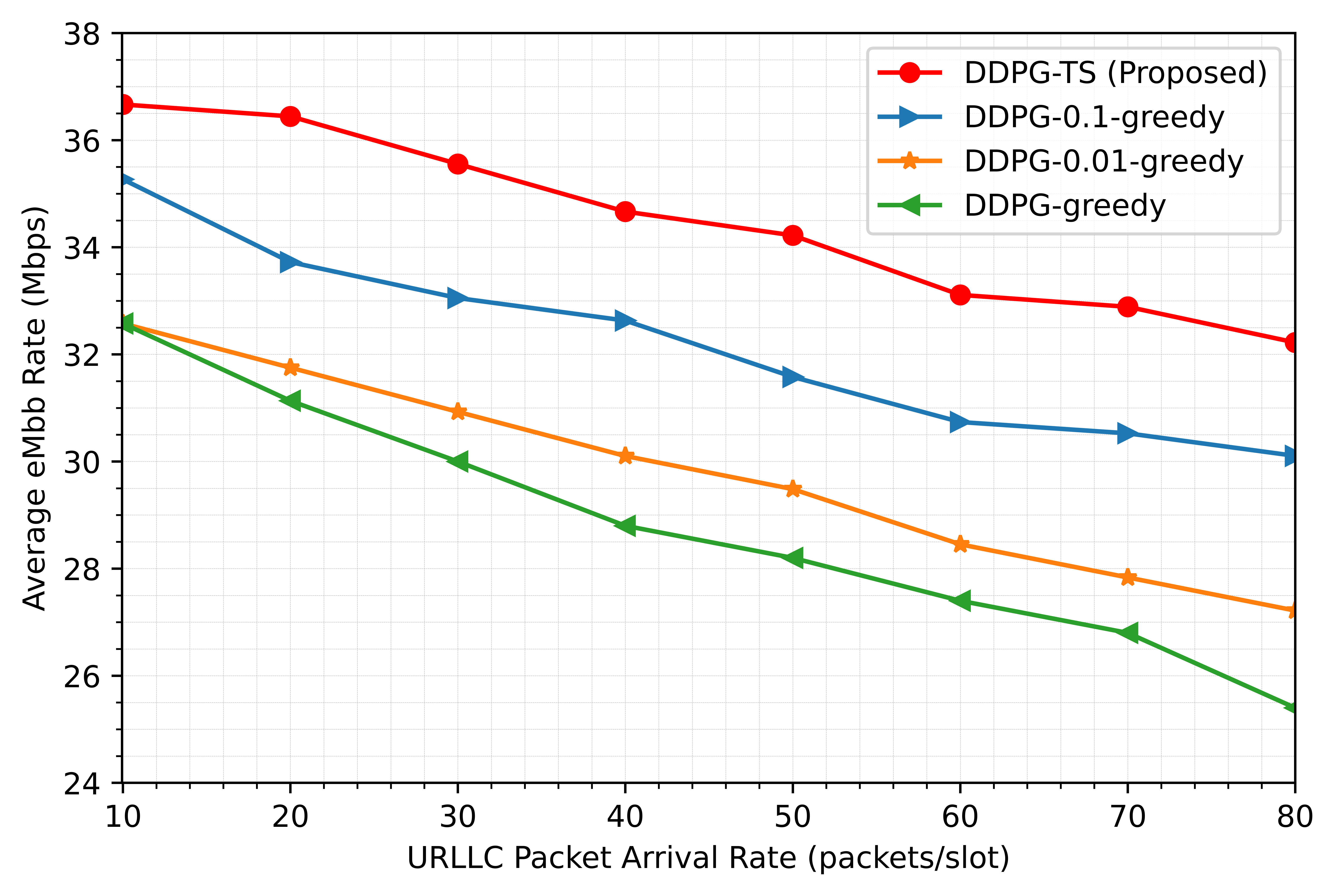}
\centering
\caption{Impact of URLLC traffic}
\label{fig:fig2}
\end{figure}
\section{Performance Evaluation}
We consider four BSs in our network model, where each BS provides coverage to an area of 250 square meters and handles eMBB users as well as URLLC users. The eMBB users generate constant full-buffer traffic, while URLLC users produce Poisson traffic with an arrival rate denoted as $\phi$. This stochastic nature reflects the sporadic demands typical in URLLC applications, where real-time responsiveness and reliability are paramount. Within this framework, the network must balance the differing requirements of eMBB and URLLC traffic, optimizing resource allocation and scheduling algorithm to meet the diverse needs of both user types.
We define the simulation parameters in Table \ref{tab:1}. 
The proposed approach undergoes training by varying the network environment. \\
\textbf{Impact on eMBB rate:}
We investigate the impact of puncturing on the eMBB data rate and conduct a comparative analysis with alternative methods across different loads of incoming URLLC traffic. 
In Fig. \ref{fig:fig2}, we illustrate the impact of incoming URLLC traffic on the eMBB rate. This influence arises from prioritizing URLLC service users, leading to the allocation of additional radio resources to meet the stringent latency requirements associated with URLLC services. Our proposed method strikes an effective balance between exploration and exploitation, allowing the algorithm to explore the solution space effectively while exploiting the knowledge gained. On the other hand, greedy-based methods where $\epsilon$ value is varied often prioritize the exploitation of known strategies rather than exploring new possibilities. This can lead to sub-optimal solutions when the environment is complex. 
However, as the influx of URLLC traffic intensifies, the average data rate begins to decline, yet the proposed approach consistently maintains a higher level compared to greedy approaches.\\
\begin{figure}[t!]
     \centering
     \begin{subfigure}[b]{0.45\textwidth}
         \includegraphics[width=\textwidth]{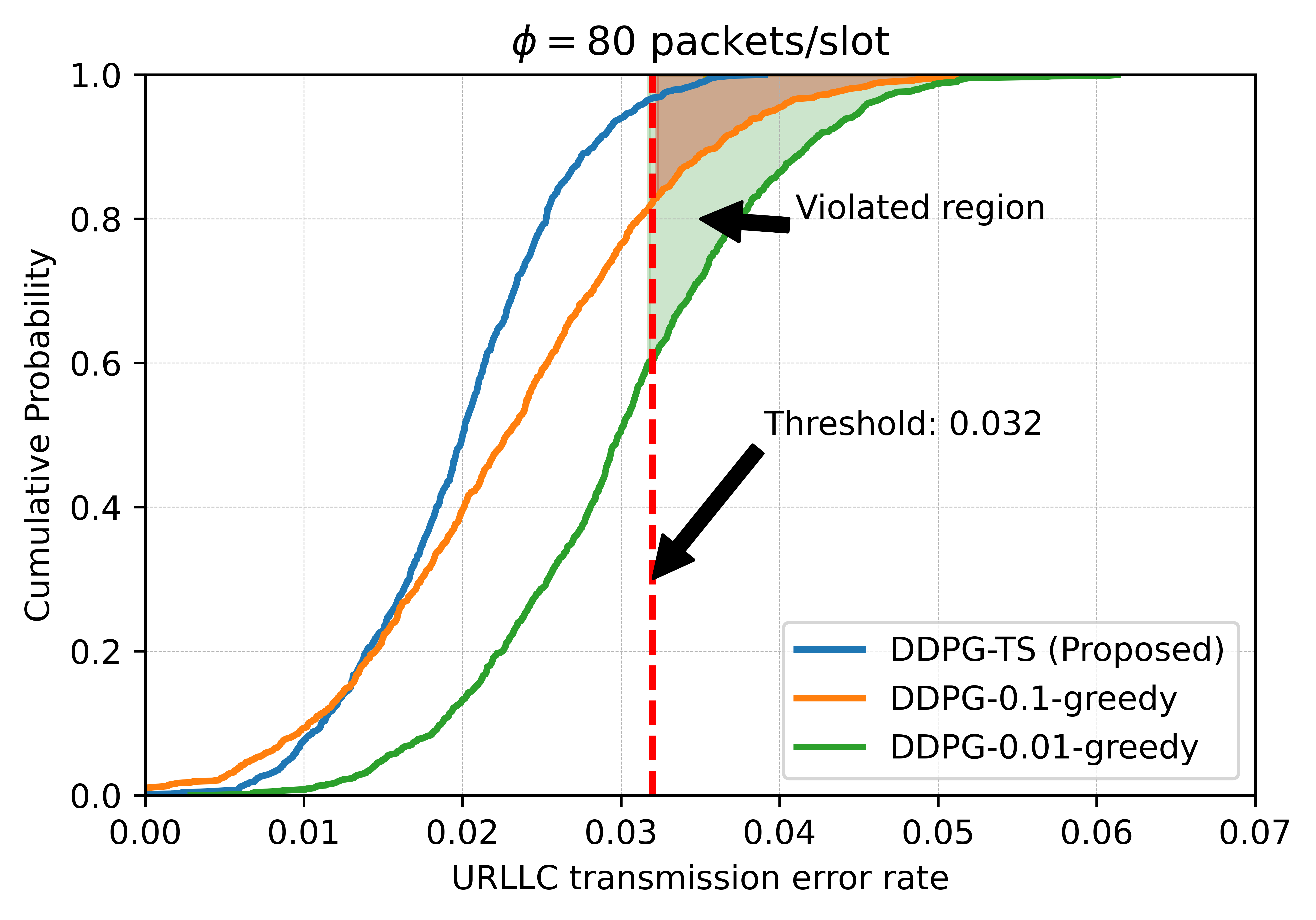}
         \caption{CDF of URLLC transmission error probability when $\phi=80$ packets/slot}
         \label{fig:fig4}
     \end{subfigure}
     \hfill
     \begin{subfigure}[b]{0.45\textwidth}
         \includegraphics[width=\textwidth]{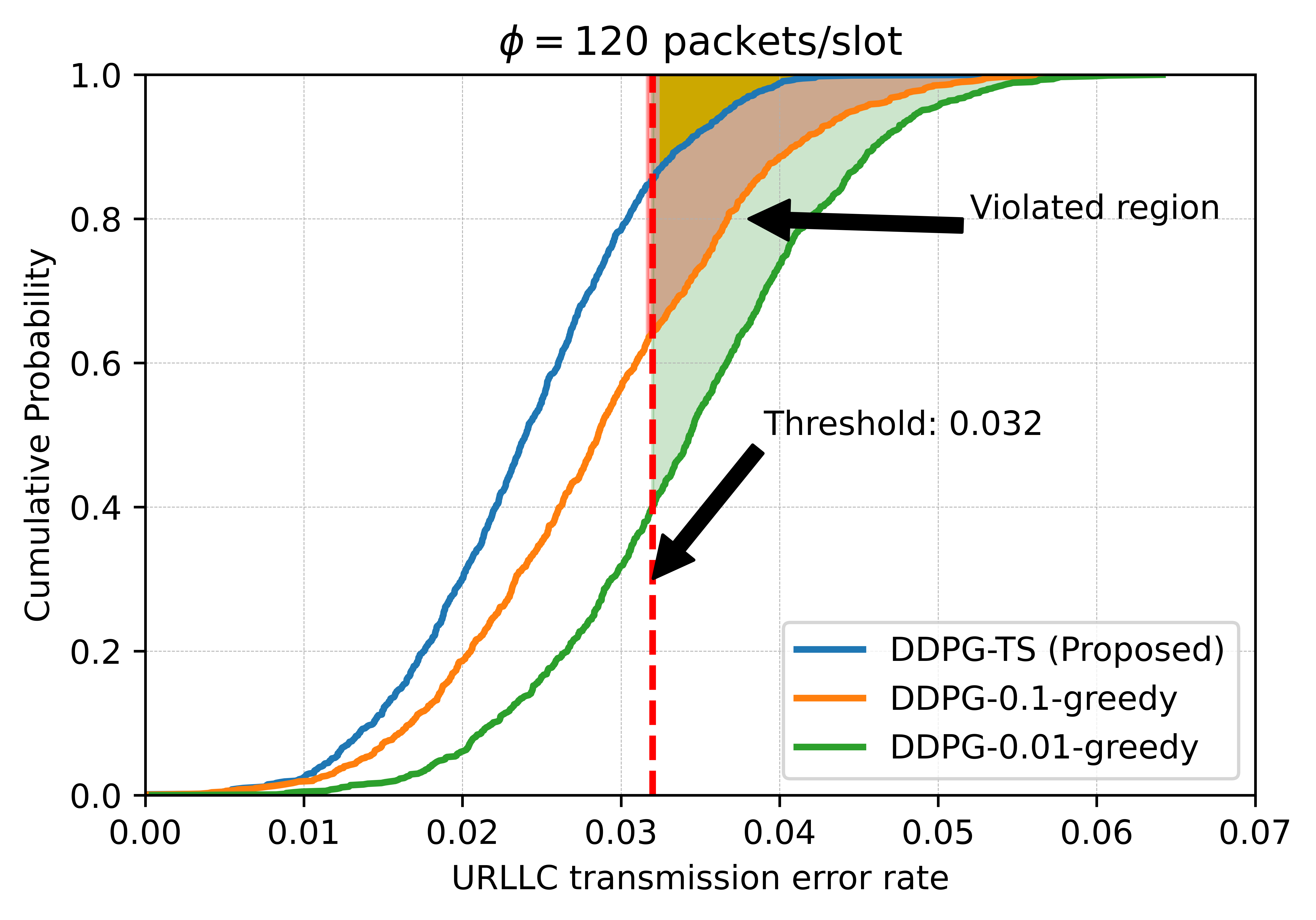}
         \caption{CDF of URLLC transmission error probability when $\phi=120$ packets/slot}
         \label{fig:fig3}
     \end{subfigure}
     \caption{CDF of URLLC transmission error probability for different $\phi$.}
\end{figure} 
\textbf{URLLC reliability:}
In Fig. 3, we examine the performance in terms of URLLC reliability by plotting the CDF of the transmission error probability. Fig. \ref{fig:fig4} shows that in more than 98\% of instances, the proposed approach ensures that the transmission error rate remains within acceptable limits. When $\phi$ is 80 packets per time-slot, the proposed method manages to uphold a high level of accuracy and reliability in transmitting URLLC packets. 
However, it is crucial to note that in Fig. \ref{fig:fig3}, increasing URLLC traffic rates can pose challenges to URLLC reliability. Our proposed method still maintains reliability in transmitting URLLC packets compared to varied $\epsilon$ values of the greedy method. 
Unlike the greedy method, which relies on fixed $\epsilon$ values and can become trapped in sub-optimal choices, Thompson sampling continually explores and exploits new strategies, ensuring that the system can respond effectively to changing traffic patterns and network states.
This adaptability is crucial for URLLC, where even minor delays or packet losses can have significant repercussions. By efficiently balancing the exploration of new transmission strategies with the exploitation of known successful ones, Thompson sampling minimizes the risk of prolonged failure modes. 
It is evident from the results that Thompson sampling improves performance by introducing a probabilistic element that helps deal with uncertainties compared to fixed learning rates.  
\section{Conclusion}
This study has addressed the resource allocation challenges in multi-cell wireless systems catering to eMBB and URLLC users. Through the formulation of an optimization problem and the development of a distributed learning framework, particularly leveraging a Thompson sampling-based DRL approach,we have presented a solution capable of making online resource allocation decisions.  
The proposed approach aligns with the evolving O-RAN network architectures, allowing for efficient learning over wireless networks. Our simulation results demonstrate the effectiveness of the algorithm in meeting the QoS requirements for both eMBB and URLLC users. This research contributes valuable insights into optimizing resource utilization in dynamic and heterogeneous wireless environments, paving the way for enhanced network performance and user satisfaction across diverse communication services.
\section*{Acknowledgment}
This work was supported by EPSRC projects, CHEDDAR EP/X040518/1 and CHEDDAR Uplift EP/Y037421/1.

\ifCLASSOPTIONcaptionsoff
  \newpage
\fi


\bibliographystyle{IEEEtran}
\bibliography{ref}

\end{document}